\documentclass{kapproc} 

\usepackage{procps}

\usepackage[dvips]{graphicx}

\upperandlowercase

 \let\footnote\savefootnote

\normallatexbib 

\begin{document}

\articletitle
{Volume Filling factors of the DIG}

\chaptitlerunninghead{Filling factor} 

\author{Dipanjan Mitra,\altaffilmark{1} Elly M. Berkhuijsen,\altaffilmark{1}
\& Peter M\"uller\altaffilmark{1}}

\affil{\altaffilmark{1}Max-Planck-Institut f\"ur Radioastronomie}



\begin{abstract}
Combining dispersion measures, distances and emission measures for 157
pulsars lying above $\mid b \mid > 5^{\circ}$ and between
$60^{\circ}<l<360^{\circ}$ we find the mean volume filling factor
($\bar{f_v}$) of the diffused ionized gas in the Milky Way.  This
filling factor is inversely related to the mean electron density
($\bar{n_c}$) in the clouds, $\bar{f_v}=(0.0184 \pm 0.0014)
\bar{n_c}^{-1.07 \pm 0.03}$, implying a nearly constant average
electron density distribution within 3 kpc of the sun.
\end{abstract}

\begin{keywords}
Interstellar medium - Milky Way
\end{keywords}

\section{Introduction}
The diffuse ionized gas (DIG) in the interstellar medium (ISM)
spreads around the dense classical HII regions and the spiral arms
with a scale height of 900 pc (Reynolds 1991). The DIG is best studied
at Galactic latitudes $\mid b \mid >5^{\circ}$, above which the
classical HII regions are largely absent. First estimates of the mean
volume filling factor ($\bar{f_{v}}$) were made by Reynolds (1977) who
derived a lower limit of $\bar{f_{v}} > 0.1$ from emission measures
(EM) and dispersion measures (DM) of 24 pulsars. Kulkarni \& Heiles
(1988) estimated the variation of $\bar{f_{v}}$ and the mean density
$\bar{n_{c}}$ in clouds with distance from the Galactic plane $z$ and
found $\bar{f_{v}}(z)$ to increase in an exponential fashion. Pynzar
(1993) was the first to find a relation between $\bar{f_{v}}$ and
$\bar{n_{c}}$, not only for the DIG but also for the classical HII
regions, $\bar{f_{v}} \propto \bar{n_c}^{-0.7}$.

We use the WHAM H$\alpha$ survey (Haffner et al. 2003 (in press)) to
estimate EM for 157 carefully selected pulsars, which we believe are
seen along lines of sight through the DIG. Using the DM of these
pulsars, and estimated distances from the model of Cordes \& Lazio
(2002), we find $\bar{f_{v}}$ and $\bar{n_{c}}$ and investigate their
relationship. We also search for variations of $\bar{f_{v}}$ and
$\bar{n_{c}}$ as a function of $z$ and determine their exponential
scale height. Details of this work can be found in Berkhuijsen, Mitra
\& M\"uller (2003, hereafter BMM).

\section[vff]
{Basic Relations}

\noindent
The dispersion measure (DM) and emission measure (EM) towards a pulsar
with distance $D$ can be written as
\begin{equation}
{\rm \frac{DM}{cm^{-3}pc}} = \int_{0}^{D} n_e(l)dl =
<n_e>D=\bar{n_c}\bar{f_v}D=\bar{n_c}L_e~,
\label{eq1}
\end{equation}

\begin{equation}
{\rm \frac{EM}{cm^{-6}pc}} = \int_{0}^{D} n_{e}^{2}(l)dl =
<n_{e}^{2}>D=\bar{n_{c}^{2}}\bar{f_v}D=\bar{n_{c}^{2}}L_e
\label{eq2}
\end{equation}
where $n_e(l)$ (in ${\rm cm^{-3}})$ is the electron density at a point
$l$ along the line of sight (LOS), $L_e$ (in pc) the total path length
through the regions containing free electrons and $\bar{n_c}$ (in
${\rm cm^{-3}}$) the average electron density in these regions which
is the mean electron density of a cloud if constant for all clouds
along the LOS. $<n_e>$ and $<n_{e}^{2}>$ are averages along $D$ and
$\bar{f_{v}} = L_e/D$ is the fraction of the LOS occupied by
electrons\footnote{ Note that the filling factor measure here is
actually a LOS filling factor, ${f_D}$, however BMM show that for
random geometrical shapes of clouds along the LOS $f_D \sim f_v$}. All
quantities with overbars are averages along the LOS. Thus combining
Eqs.(\ref{eq1}) and (\ref{eq2}) we find,
\begin{equation}
\bar{f_{v}} = \frac{L_e}{D} =
\frac{<n_e>}{\bar{n_c}}=\frac{<n_{e}^{2}>}{{\bar{n_{c}^{2}}}}~.
\label{eq3}
\end{equation}
\noindent
Other useful expressions are:
\begin{equation}
\bar{n_c} = \frac{{\rm DM}}{L_e}, L_e = \frac{{\rm DM^2}}{{\rm
EM}}~~{\rm and}~\bar{f_v}=\frac{{\rm DM^2}}{{\rm EM}~D}~.
\label{eq4}
\end{equation}

\begin{figure}[ht]
\centerline{\includegraphics[width=5 cm,height=
10cm,angle=-90]{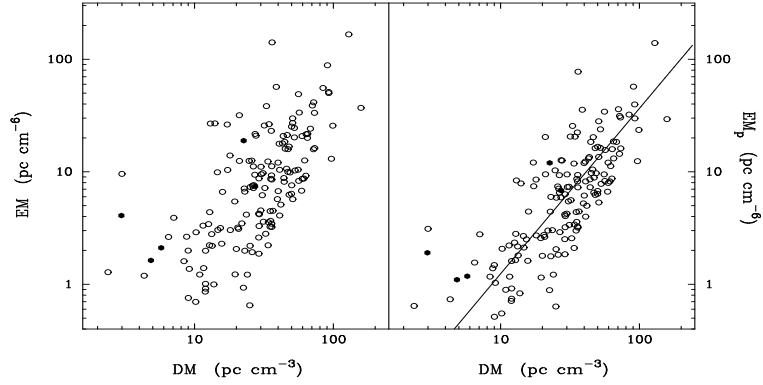}}
\caption{The uncorrected (left) and corrected (right) EM-DM relations are shown.
The full line in the right-hand plot shows the bisector fit to the data.}
\label{fig1}
\end{figure}

\begin{figure}[ht]
\centerline{\includegraphics[width=3.5 cm,height= 6cm,angle=-90]{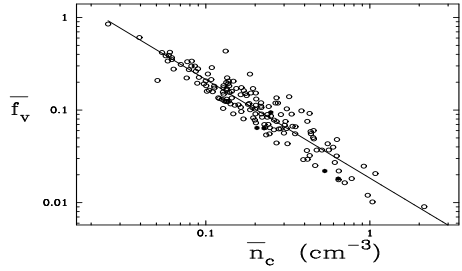}}
\caption{The variation of $\bar{f_v}$ versus $\bar{n_c}$ is shown.
The full line is a power law fit to the data.}
\label{fig2}
\end{figure}

\begin{figure}[ht]
\centerline{\includegraphics[width=5 cm,height= 8cm,angle=-90]{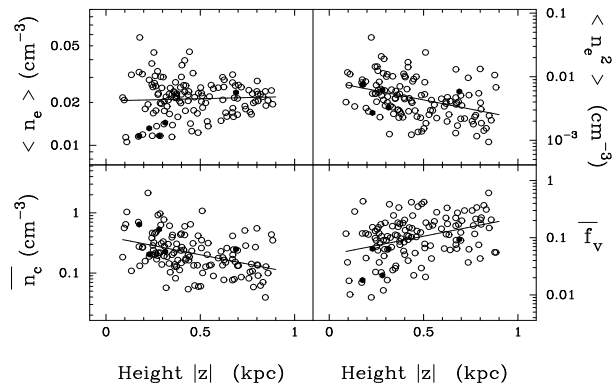}}
\caption{Variations of $<n_e>$, $<n_e^{2}>$, $\bar{n_c}$ and $\bar{f_v}$ with
$z$. The full lines are the exponential fits to the data.}
\label{fig3}
\end{figure}

\section{Data and Results}\inxx{Making tables}
To evaluate $\bar{n_{c}}$ and $\bar{f_v}$ from Eqs.(\ref{eq1}) to
(\ref{eq4}), EM, DM and $D$ are needed. To obtain EM we looked into
the recently available WHAM H$\alpha$ survey and found EM for 744
pulsars in the overlapping survey region. However, in several
directions HII regions along the LOS can affect our analysis.  To
reduce their influence we chose pulsars lying at $\mid b \mid >
5^{\circ}$ and excluded several `anomalous' directions above the plane
(see BMM for details) leaving 157 pulsars for our analysis.  We took
DM from the ATNF pulsar
catalogue\footnote{http://www.atnf.csiro.au/research/pulsar/psrcat/}
and obtained pulsar distance $D$ from the recent electron density
model of Cordes \& Lazio (2002).

EM was corrected for absorption of H$\alpha$ emission by dust along
the LOS as ${\rm EM_{c} = EM} e^{(0.086\pm0.005/sin\mid b \mid )}$
(Diplas \& Savage 1994).  ${\rm EM_c}$ was corrected for the extra
emission coming from beyond the pulsar to obtain emission measure
along the LOS to the pulsar as ${\rm EM_p = EM_c} (1-e^{-\mid z_p \mid
/h})$, where $z_p$ is the perpendicular distance to the pulsar from
the Galactic plane and $h=280$ pc is the exponential scale height of
$n_e^{2}$ (see BMM for details).

\vspace{0.2cm}
Using these data we obtained the following power-law
fits:

the EM${\rm _p}-$DM relation: EM${\rm _p} = (0.042 \pm 0.014)~{\rm
DM}^{1.47 \pm 0.09}$ [see Fig.(\ref{fig1})]

the $\bar{f_{v}} - \bar{n_c}$ relation: $\bar{f_v} = (0.0184 \pm
0.0011)~\bar{n_c}^{-1.07 \pm 0.03}$ [see Fig.(\ref{fig2})]
\vspace{0.2cm}

The variations of several quantities with z (in kpc) were fitted by
exponentials [as shown in Fig.(\ref{fig3})]:

$<n_e> = (0.0205\pm0.0014)~ {\rm{exp}}({\mid z \mid / 14^{}_{-8}})$

$<n_{e}^{2}> = (0.0084\pm0.0012)~ {\rm{exp}}({-\mid z \mid /
0.75^{+0.20}_{-0.13}})$

$\bar{n_{c}} = (0.407\pm0.059)~ {\rm{exp}}({-\mid z \mid /
0.71^{+0.18}_{-0.12}})$

$\bar{f_{v}} = (0.0504\pm0.0095)~ {\rm{exp}}({\mid z \mid /
0.67^{+0.20}_{-0.13}})$
\vspace{0.2cm}

Here we point out two important effects characterizing the DIG.  1.)
The remarkable correlation and inverse relation of $\bar{f_{v}}$ with
$\bar{n_c}$ is due to the near constancy of $<n_e>$.  This suggests
that the DIG is in thermal pressure equilibrium and/or has a turbulent
fractal structure.  2.) The scale height of $\bar{f_{v}}$ is about 0.7
kpc and the mean size of the clouds increases at larger Galactic
heights. The physical reasons for these relationships need to be
investigated.

\begin{acknowledgments}
DM wishes to thank the organizing commitee for their financial
support, given to attend the conference. The Wisconsin H-Alpha Mapper
is funded by the National Science Foundation.
\end{acknowledgments}

\end{document}